\documentstyle[aps,prl,epsfig,amsfonts]{revtex}
\newcommand{\be}{\begin{equation}}
\newcommand{\ee}{\end{equation}}
\newcommand{\bea}{\begin{eqnarray}}
\newcommand{\eea}{\end{eqnarray}}

\newcommand{\lb}{\label}
\newcommand{\ind}[2]{^{#1}_{\mbox{\scriptsize #2}}}

\begin{document}
\pagestyle{myheadings}
\parindent 0mm
\parskip 6pt

\centerline{\Large{\bf Bound state approach to the QCD coupling }}
\centerline{\Large{\bf at low energy scales}}

\vspace{0.5truecm}
\centerline{\bf{M.\ Baldicchi$^{\dagger}$, A.\ V.\ Nesterenko$^{\ast}$,
G.\ M.\ Prosperi$^{\dagger}$,}}
\centerline{\bf{D.\ V.\ Shirkov$^{\ast}$ and C.\ Simolo$^{\dagger}$}}
\vspace{0.3truecm}
{}\centerline{\small{$^{\dagger}\,$ Dip. di Fisica, Universit\`a
di Milano and INFN, Sezione di Milano }}
\centerline{\small{Via Celoria 16, 20133 Milano, Italy}}
{}\centerline{\small{$^{\ast}\,$ Bogoliubov Laboratory of Theoretical
Physics}}
{}\centerline{\small{
Joint Institute for Nuclear Research, Dubna, 141980, Russia}}

\begin{abstract}
\noindent
We exploit theoretical results on the meson spectrum within the
framework of a Bethe-Salpeter (BS) formalism adjusted for QCD, in order to
extract an ``experimental'' coupling $\alpha_{s}^{\rm exp}(Q^2)$
below 1~GeV by comparison with the data. 
Our results for $\alpha_{s}^{\rm exp}(Q^2)$ exhibit a
good agreement with the infrared safe Analytic Perturbation Theory
(APT) coupling from 1~GeV down to 200~MeV.\\ 
As a main result, we claim that the combined BS-APT theoretical
scheme provides us with a rather satisfactory correlated
understanding of very high and low energy phenomena.
\end{abstract}

\vspace{0.1truecm}
{\bf I.Introduction}. 
The Renormalization Group (RG) improved perturbative QCD yields a
consistent picture of high energy strong interaction processes from
a few GeV up to a few hundred GeV scale~\cite{Bethke:2006ac}.
At the same time, in the low energy domain the very existence of the
unphysical (so-called ``Landau'') singularities in both the
RG-invariant coupling $\alpha_s(Q^2)$ and physical observables
contradicts the general principles of the local QFT and spoils the
theoretical analysis of low energy hadron dynamics. In
particular, the ghost-pole issue gives rise to severe complications
as far as the bound states problem is concerned, since the scale $Q$
(i.e., the momentum transfer in the $q\bar q$ interaction)
involved is typically below 1~GeV. Moreover,
results of lattice simulations testify to the absence of spurious IR
singularities in the QCD coupling~\cite{dvFourier03,Lattice}.
A reliable algorithm to get rid of these singularities is
provided by the APT approach~\cite{ShSol96-7}, based on the 
causality condition 
which imposes $\alpha_s(Q^2)\,$ to satisfy a dispersion
relation with only the physical cut $- \infty < Q^2 <0\,$.\\
This prescription has been exploited in the framework of a second
order Bethe-Salpeter (BS) like formalism \cite{BMP} for the
calculation of the meson spectrum in the light and heavy quark
sectors.
The model is derived from the QCD Lagrangian taking
advantage of a Feynman-Schwinger representation for the solution
of the iterated Dirac equation in an external field. Confinement is
encoded through an ansatz on the Wilson loop correlator; indeed
the quantity $i\ln W$ is written as the sum of a one-gluon
exchange~(OGE) and an area term
\begin{equation}
 i\ln W = (i\ln W)_{\rm OGE}+ \sigma S \,.
\label{eq:wilson}
\end{equation}
By means of a three dimensional reduction, the original BS
equation takes the form of the eigenvalue equation for 
\begin{equation}
    M^2 = M_0^2 + U_{\rm OGE}+U_{\rm Conf} \,,
\label{eq:M2}
\end{equation}
where $M^2$ is the squared bound state mass, 
$ M_0 = w_1+w_2 = \sqrt{m_{1}^2 + {\bf k}^2} + \sqrt{m_{2}^2 + {\bf
     k}^2}$,
${\bf k}$ the c.m.\ momentum of quarks, $m_1$ and $m_2$ their
constituent masses and $U=U_{\rm OGE}+U_{\rm Conf}$ the resulting potential.
The combined BS-APT theoretical scheme was clearly supported by the results
of previous computations performed in~\cite{BP,Baldicchi:2004wj}
by using a 1-loop APT coupling
$\alpha^{(1)}_E(Q^2)$, with an effective scale constant
$\Lambda_{n_f=3}^{(1,{\rm eff})}\simeq 200\,$MeV (see Eq.\
(\ref{ARC1L}) below). A substantial agreement of the spin
averaged c.o.g.\ masses with the data is achieved throughout the whole
spectrum and the splittings
$1^3S_1$-$1^1S_0$ well reproduced in all sectors involving
light, strange and charm quarks. 
Among other attempts to study meson properties,
by taking relativistic effects into account, we remind 
e.g.~\cite{Faustov} and Refs. therein. They differ in the ansatz by
which confinement is introduced and in the method used in the
determination of bound states: 
quasipotential~\cite{Faustov,Simonov,Roberts,Sazdjian}, Green function
\cite{Simonov} and first order BS formalism \cite{Roberts,Sazdjian}.\\
In this note we summarize the main results of an investigation
performed from the reversed point of view, i.e. by 
exploiting the results on the meson spectrum within the BS approach
in order to extract ``experimental'' QCD coupling
$\alpha_s^{\rm exp}(Q^2)$ below 1~GeV, by
comparison with meson mass data. 
The results are twofold. On the one hand, the 3-loop
APT coupling reasonably fits $\alpha_{s}^{\rm exp}(Q^2)$ from
1~GeV down to 200~MeV, quantitatively confirming the relevance of
the APT approach to IR phenomena. On the other hand,
below this scale, the experimental points give a slight hint
about the vanishing of $\alpha_{s}(Q^2)\,$,  
or the existence of a finite limit lower than $1/\beta_0\,$,
as $Q\to0$. This could correlate 
with some results from lattice simulations~\cite{Lattice}
and can be theoretically discussed in
the framework of a recent ``massive'' modification~\cite{MAPT} of
APT. Note that a non vanishing freezing value as suggested 
in~\cite{Badalian:2001} would be still consistent with our results
for $ Q < 200 $ MeV, but does not agree with the general trend of 
our results in the region 200-500~MeV. For the detailed set of our results and 
technicalities we refer to the extensive account~\cite{BNPS}. 

\vspace{0.1truecm}
{\bf II.The causal APT coupling}.
A number of non-perturbative tricks to handle the ghost-pole
problem was reviewed in~\cite{Prosperi:2006hx}. Here we exploit 
the APT approach to QCD (see~\cite{ShSol96-7,APT-07}), in which  
the RG-improved power series in 
$\alpha_s(Q^2)$ for a given ``Euclidean'' observable is replaced 
by a non power expansion over the set of functions
\begin{equation}
\label{AAPT}
{\mathcal A}_{n}(Q^2) = \int\limits_{0}^{\infty}
\frac{\rho_{n}(\sigma)}{\sigma + Q^2}\, d\sigma\,;\quad
\rho_{n}(\sigma) = \frac{1}{\pi}\,\mbox{Im}\!
\left[\alpha_s(-\sigma - i \varepsilon)\right]^{n}\,.
\end{equation}
Here the first function $\alpha_E(Q^2)\equiv{\mathcal A}_1(Q^2)$
plays the
role of the APT Euclidean coupling, and at 1-loop
it reads
\begin{equation}
\label{ARC1L}
\alpha_{E}^{(1)}(Q^2) = \frac{1}{\beta_{0}}\!\left[\frac{1}
{\ln(Q^2/\Lambda^2)}+\frac{\Lambda^2}{\Lambda^2-Q^2}\right]\,.
\end{equation}
At the higher loops Eq.~(\ref{AAPT}) with $n=1$ can be integrated
only numerically (for details see \cite{Kurashev:2003pt}). Nevertheless, for
practical applications below 1~GeV one can resort to the same 
Eq.~(\ref{ARC1L}) with modified scale constant 
(see Refs.~\cite{BNPS,APTApprox}). It is relevant to the problem in hand
to mention a recently devised ``massive''
modification for the QCD analytic charge~\cite{MAPT}. The point is
that the representation~(\ref{AAPT}) does not hold for every QCD
quantity, and the effect of a non vanishing mass threshold~$m$ in the
dispersion relations could play a substantial role. Then the set
of the APT functions~(\ref{AAPT}) should be replaced by the
set of the ``massive'' ones with an adjustable parameter~$m$ 
\begin{equation}
\label{AMAPT}
\textsf{A}_{n}(Q^2, m^2) = \frac{Q^2}{Q^2 + 4 m^2}
\int\limits_{4 m^2}^{\infty} \rho_{n}(\sigma)\, \frac{\sigma -
4 m^2}{\sigma + Q^2}\, \frac{d\sigma}{\sigma}\,.
\end{equation}
The first function still plays the
role of the ``massive'' coupling $\alpha(Q^2, m^2)$, with universal limit 
$\alpha(0, m^2)=0\,$. 

\vspace{0.1truecm}
{\bf III.The quark-antiquark spectrum}.
Similarly to Refs.~\cite{BP,Baldicchi:2004wj}, we
neglect the spin-orbit and tensorial like terms in both the  
perturbative and the confining part of BS potential,
$U_{\rm OGE}\,$ and $U_{\rm Conf}\,$. Among the spin
dependent terms only the hyperfine splitting one
proportional to ${1\over 6} {\bf \sigma}_1 \cdot {\bf \sigma}_2\,$
is retained. Then one has
\be
\langle {\bf k} \vert U_{\rm OGE}  \vert {\bf k}^\prime \rangle
= {4\over 3} {\alpha_{s}({\bf Q}^2) \over \pi^2}
     \rho({\bf k},{\bf k'})
     \Bigg[ - \frac{1}{{\bf Q}^2}
     \bigg( q_{10} q_{20} + {\bf q}^2 - { ( {\bf Q}\cdot {\bf q})^2 \over
{\bf Q}^2 } \bigg)
 + {1\over 6} {\bf \sigma}_1 \cdot {\bf \sigma}_2  \Bigg ]\,, \quad
\lb{upt}
\ee
\be \langle {\bf k} \vert
U_{\rm Conf}  \vert {\bf k}^\prime \rangle ={\sigma\over ( 2
\pi)^3}  \rho({\bf k},{\bf k'}) \int d^3{\bf r}\,
     e^{i {\bf Q}\cdot{\bf r}}
J^{\rm inst}({\bf r}, {\bf q}, q_{10}, q_{20})\,,
\lb{ucf}
\ee
\bea
&& J^{\rm inst}({\bf r}, {\bf q}, q_{10}, q_{20})= { r \over
q_{10}+q_{20}}
   \left[ q_{20}^2 \sqrt{q_{10}^2-{\bf q}^2_\perp} +
q_{10}^2 \sqrt{q_{20}^2 - {\bf q}_\perp^2}\right. +\qquad\nonumber\\
& & \qquad\left. + {q_{10}^2 q_{20}^2 \over \vert
{\bf q}_\perp \vert}
   \left(\arcsin{\vert{\bf q}_\perp \vert \over q_{10} }
   + \arcsin{\vert {\bf q}_\perp\vert \over q_{20}}\right)\right]\,.
   \label{eq:uconf1}
\eea
Here, $\,{\bf q}={{\bf k}+{\bf k}^\prime\over 2}\,,
\,{\bf Q}={\bf k}-{\bf k}^\prime\,,\,q_{j0}={w_j+w_j^\prime\over 2}\,$,
$w_j = \sqrt{m_{j}^2 + {\bf k}^2}$,
and $\rho({\bf k},{\bf k'})=\sqrt{(w_1+w_2) (w_1^\prime + w_2^\prime)
\over w_1 w_2 w_1^\prime w_2^\prime}\,$, while $m_{1}$ and $m_{2}$
denote the constituent quark and antiquark masses. 
For the complete expression of the potential and technical details 
we refer to~\cite{BMP,Baldicchi:2004wj} and~\cite{BNPS}.
The meson masses have been computed by the equation
\be m^2_{a}
=\langle\phi_a|M_0^2|\phi_a\rangle+ \langle\phi_a|U_{\rm
OGE}|\phi_a\rangle+ \langle\phi_a|U_{\rm Conf}|\phi_a\rangle\,,
\lb{m_th}
\ee
where $\phi_a$ is a zero-order wave
function for the state $a$ ($a$ being the whole set of quantum
numbers), obtained by solving the eigenvalue equation for the
static limit Hamiltonian $H_{\rm CM} = w_1 + w_2 - {4 \over 3}
{\alpha_{s} \over r } +  \sigma r \,$ by the Rayleigh-Ritz
method. To this a second order correction in
the hyperfine term was added in some cases. 
The hurdle of spurious singularities has been avoided by replacing
$\alpha_s(Q^2)$ in~(\ref{upt})  with $\alpha_E^{(1)}(Q^2)$ as given
by (\ref{ARC1L}) with an effective QCD scale
$\Lambda_{n_f=3}^{(1, \rm{eff})}=193\,$MeV.
This value has been chosen by imposing that $\alpha_E^{(1)}(Q^2)$
crosses the 3-loop APT coupling $\alpha_E^{(3)}(Q^2)$ at $Q=0.65\,$GeV, where  
$\alpha_E^{(3)}(Q^2)$ is normalized along with the world 
average~\cite{pdg} $\alpha_s(M_Z^2)=0.1176(20)\,$, corresponding
to $\Lambda_{n_f=5}^{(3)}=236\,$MeV and $\Lambda_{n_f=3}^{(3)}=417\,$MeV 
by continuous threshold matching. 
The relative difference between
the 1-loop effective and 3-loop exact APT curves is no more than 
2$\%$ in the region $0.4<Q<1.0\,$GeV, to which the bulk of effective 
$Q_a$ belongs, and it is enhanced up to 7$\%$ only at $Q\sim 0.2\,$GeV. 
The string tension has been fixed a priori to the value
$\sigma=0.18\,\,{\rm GeV^2}\,$ consistently with lattice
simulations.
The light and heavy quark masses are then determined by fitting the $\pi\,$,
$\phi\,$, $J/\psi\,$ and $\Upsilon\,$ masses~\cite{pdg}. It turns out 
$m_u=m_d=196\,$MeV, $m_s=352\,$MeV, $m_c=1.516\,$GeV and
$m_b=4.854\,$GeV. 
Within this framework an overall agreement with experimental data is
a\-chie\-ved throughout the spectrum (see~\cite{BNPS} for the complete
set of results, preliminary results were given
in~\cite{Baldicchi:2006az}).

\vspace{0.1truecm}
{\bf IV.Extracting $\alpha_{s}^{\rm exp}(Q^2)$ from the data}.
As stated, we focus our attention here on the reversed problem, i.e., the
determination of $\alpha_{s}$ at the characteristic scales of a
selected number of ground and excited states. 
As a first step, we associate with each state $a$ an effective momentum 
transfer $Q_a$ defined by the relation
\be \langle\phi_a|U_{\rm OGE}|\phi_a\rangle
\equiv\langle\phi_a|\alpha_{E}^{(1)}({\bf Q} ^2){\mathcal O}
({\bf  q};{\bf Q}) |\phi_a\rangle=\alpha_{E}^{(1)}(Q_a^2)
\langle\phi_a|{\mathcal O} ({\bf q};{\bf Q}) |\phi_a\rangle\,,
\lb{a_th}
\ee
i.e. as the value of $Q$ for which the fixed coupling value 
$\alpha_{E}^{(1)}(Q_a^2)$ inserted in (\ref{m_th}) reproduces the
same mass $m_a$ as when using the running coupling $\alpha_{E}^{(1)}(Q^2)\,$.
The quantity ${\mathcal O}({\bf q};{\bf Q})$ in Eq.~(\ref{a_th}) can be drawn
by the second line of Eq.~(\ref{upt}). 
Then, the experimental coupling $\alpha_{s}^{\rm exp}(Q_a^2)$
can be defined by
\be
\langle\phi_a|M_0^2|\phi_a\rangle+ \alpha_{s}^{\rm
exp}(Q^2_a)\langle\phi_a|{\mathcal O}({\bf q};{\bf Q})
|\phi_a\rangle + \langle\phi_a|U_{\rm Conf}|\phi_a\rangle=m^2_{\rm
exp}\,, \lb{m_exp} 
\ee 
or, by combining Eqs.\ (\ref{m_th}), (\ref{a_th}) and (\ref{m_exp}), 
\be \alpha_{s}^{\rm exp}(Q^2_a)=\alpha_{E}^{(1)}(Q_a^2)+
\frac{m^2_{\rm exp}-m^2_{a}}
{\langle\phi_a|{\mathcal O}({\bf q};{\bf Q})|\phi_a\rangle}\,.
\lb{a_exp}
\ee
The sensitivity of
the effective $Q$'s, derived as above, has been
checked by analyzing their deviations for a $25\%\,$ shift of
$\Lambda_{n_f=3}^{(1,{\rm eff})}$ in (\ref{ARC1L}) and the 
average change in the momentum scale is about~$3\%\,$.
This shows that $\alpha_{s}^{\rm exp}(Q_a^2)$ is rather insensible to the 
specific form of $\alpha_{E}^{(1)}(Q^2)\,$, and 
justifies the use of $\alpha_{s}^{\rm exp}(Q_a^2)$ in~(\ref{m_exp}).
Obviously the theoretical meson masses are sensitive to a variation 
of the quark masses (particularly in the case of the $\pi$), 
while $\alpha_s^{\rm exp}\,$ and the relative 
$Q_a$ turn out to be much more stable. For instance,
an increase in the light quark mass of 5$\%$ amounts to a change of about 
$2\%$ in the value of $\alpha_s^{\rm exp}\,$ and $0.2\%$ in the relative
$Q_a\,$.\\ 
Note that the APT coupling, involved 
in the calculation of the spectrum, is remarkably
stable with respect to both the choice of renormalization scheme  
and the higher loop corrections~\cite{ShSo1998} ($\alpha_{E}^{(2)}(Q^2)$ 
differs from $\alpha_{E}^{(3)}(Q^2)$ by non more than 0.3~$\%$ below 600~MeV).
This makes the method essentially RS independent.\\
Our results are unavoidably model dependent due to 
ansatz~(\ref{eq:wilson}), which consists of 
the sum of two contributions that one knows to be asymptotically correct for
small and large quark-antiquark distances. More sophisticated ansatz also 
exist (see e.g.~\cite{Billo:2006zg}, \cite{Dosch:1987sk} or 
\cite{Baker:1994nq}), albeit difficult to implement
within BS formalism. 
In the context of our model the sources of error are the istantaneous
approximation implied by the three-dimensional reduction 
of the BS equation, the approximation introduced into the resolution of
the eigenvalue equation, the inclusion of only the leading perturbative
contribution in the BS kernel $I$, and finally having neglected coupling
between different
quark antiquark channels.
The NLO contribution $\Delta I$ originates essentially
from three diagrams with
two-gluon exchange; two triangular graphs with a four-line vertex
$g^2\phi^*\phi A_{\mu}A^{\mu}$ and two three-line vertices
$g\phi^*\partial_{\mu}\phi A^{\mu}\,$, and a crossing box with
four three-line vertices (Fig.~\ref{nlo}). A somewhat crude estimate of these
contributions finally yields a global error on the potential
$\Delta{\mathcal O}/{\mathcal O}\sim\Delta I/ I$, which spans
from $20\%$ for the light-light quark system to about $1\%$ for the
$ b \bar{b} $ system.
As to the last type, within
this approximation the BS masses are expected to match the experimental
ones within the half width of the state, i.e., $\Delta m_a\sim
\Gamma_a/2\,$. Keeping in mind Eq.~(\ref{a_exp}), the estimated
theoretical errors read
\begin{equation}
\Delta_{\rm NLO}\alpha_{s}\sim\alpha_{E}^{(1)}(Q_a^2)\,{\Delta
I\over I}\,, \qquad \qquad \Delta_{\rm \Gamma}\alpha_{s}=
\frac{m_a} {|\langle\phi_a|{\mathcal
O}({\bf q};{\bf Q})|\phi_a\rangle|}\,\Gamma_a\,.
\label{err}
\end{equation}
The experimental error $\Delta m_{\rm exp}$ is generally much
smaller than $\Gamma_a/2\,$. When, however, this is not the case
one must also consider the experimental error $\Delta_{\rm
exp}\alpha_{s}\,$, obtained from the second of (\ref{err}) by replacing
$m_a\,\Gamma_a$ with $2m_{\rm exp}\Delta m_{\rm exp}\,$. 
Quark self-energy effects have been taken into account by a recursive 
resolution of the Dyson-Schwinger equation. This simply amounts in our approximation
to replacing the current quark masses with 
the constituent masses~\cite{Baldicchi:2004wj}. 
All other sources of errors above mentioned 
(including model dependence), though difficult to estimate, 
can be globally taken into account by an additional
overall error $\overline{\Delta}m $
on the masses, independent of $a$, and  
chosen such that
$ \chi^{2}_{m} = 
\frac{1}{N_{SP}} \sum_{a = 1}^{N_{SP}}
(m_{a} - m_{\rm exp} )^{2}/[(\Delta_{\rm tot} m_{a})^{2}
+ (\overline{\Delta}m)^{2}]\sim 1\,$. Here 
$\Delta_{\rm tot} m_{a} $ is the total
error resulting from all sources explicitly evaluated,
$ ( \Delta_{\rm tot} m_{a} )^{2} = m_{a}^{2} \Delta I/I +
(\Gamma_{a}/2)^{2} + ( \Delta m_{\rm exp} )^{2} $
and the sum restricted to the safer S and P states.
We find $ \overline{\Delta }m \sim 20 \, {\rm MeV} $.

\vspace{0.1truecm}
{\bf V.Conclusions.} 
All results are displayed pictorially in Fig.~\ref{low}. 
Values of $\alpha_{s}^{\rm exp}$ at the same $Q$ from triplet and 
singlet states have been
combined by means of a weighted average according to their errors.
The points $\alpha_{s}^{\rm exp}(Q^2)$ show a noticeable evolution
from 500 down to 200~MeV, where only the safer S and P states are involved, 
in remarkable agreement with the 3-loop APT coupling $\alpha^{(3)}_{E}(Q^2)$ 
properly normalized.
Precisely we find
$ \chi^{2}_{\alpha} = \frac{1}{N_{SP}} \sum_{a = 1}^{N_{SP}}
( \alpha^{\rm exp}_{s}(Q_{a}^{2}) - \alpha_{E}^{(3)}(Q_{a}^{2}))^{2}/
[(\Delta_{\rm tot} \alpha_{s} )^{2} +
( \overline{\Delta} \alpha_{s} )^{2} ] \sim 0.8 $,
with 
$ \Delta_{\rm tot} \alpha_{s}$
the total error explicitly evaluated and
$ \overline{\Delta} \alpha_{s}$ 
the uncertainty obtained from the second of
(\ref{err}) by replacing $ \Gamma_{a}$ by $2\overline{\Delta} m $.
The agreement quantitatively supports the APT approach to
IR phenomena down to a few hundred~MeV.
Below 200 MeV, the experimental points exhibit a
tendency to deviate from the APT curve and
to approach zero or at least a finite limit lower the universal APT
freezing value. 
Note that these points have been obtained from high orbital excitations
(D and F states), both experimentally and theoretically
much more uncertain, as clearly shown by their large error bars.
However, the extracted points $\alpha_{s}^{\rm exp}(Q^2)$ could 
correlate with some lattice results~\cite{Lattice}, and discussed
in the framework of the ``massive'' modification~\cite{MAPT} 
of the APT algorithm (see Sec.~1) which takes into account effects of a finite
threshold.
Finally we stress that a synthesis of results for
$\alpha_{s}(Q^2)$ as defined from bound states in the BS framework
with high energy data shows a very good agreement 
with the 3-loop APT coupling. The perturbative 3-loop
coupling with IR singular behavior is ruled out by the data, 
whereas the BS-APT theoretical scheme allows a rather
satisfactory correlated understanding of very high and low
energy phenomena.

\vspace{0.1truecm}
{\bf Acknowledgments}.
The partial support of grants RFBR 05-01-00992, NS-5362.2006.2, and
BelRFBR F06D-002 is acknowledged. One of the authors (D.Sh.) would like
to thank Drs.~R.~Faustov and O.~Galkin for useful discussions, and
Dr.~R.~Pasechnik for some numerical estimates.

\newpage

\begin{figure}[!t]
\centerline{\includegraphics[width=0.7\textwidth,height=0.7\textheight]{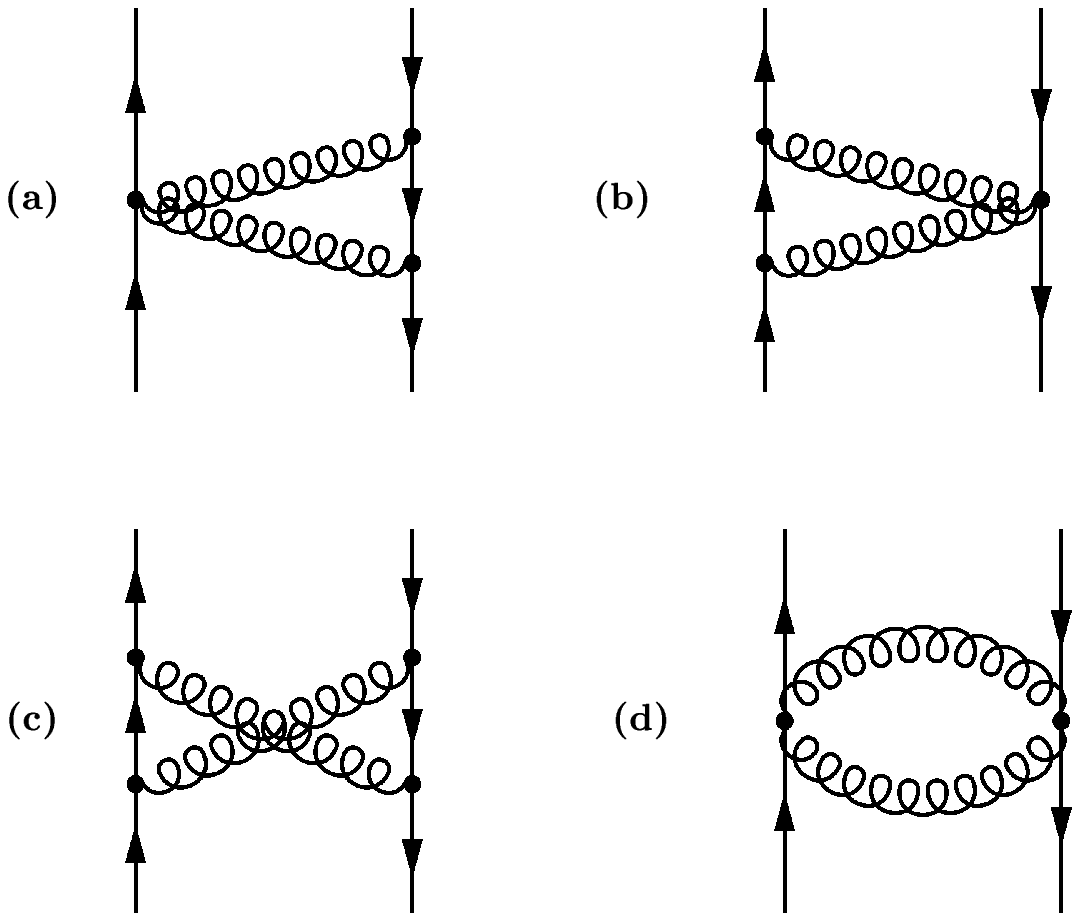}
}\caption{{\footnotesize NLO contributions to the second order BS kernel $I$.}} 
\lb{nlo}
\end{figure}


\newpage

\begin{figure}[!t]
\includegraphics[width=0.9\textwidth,height=0.55\textheight]{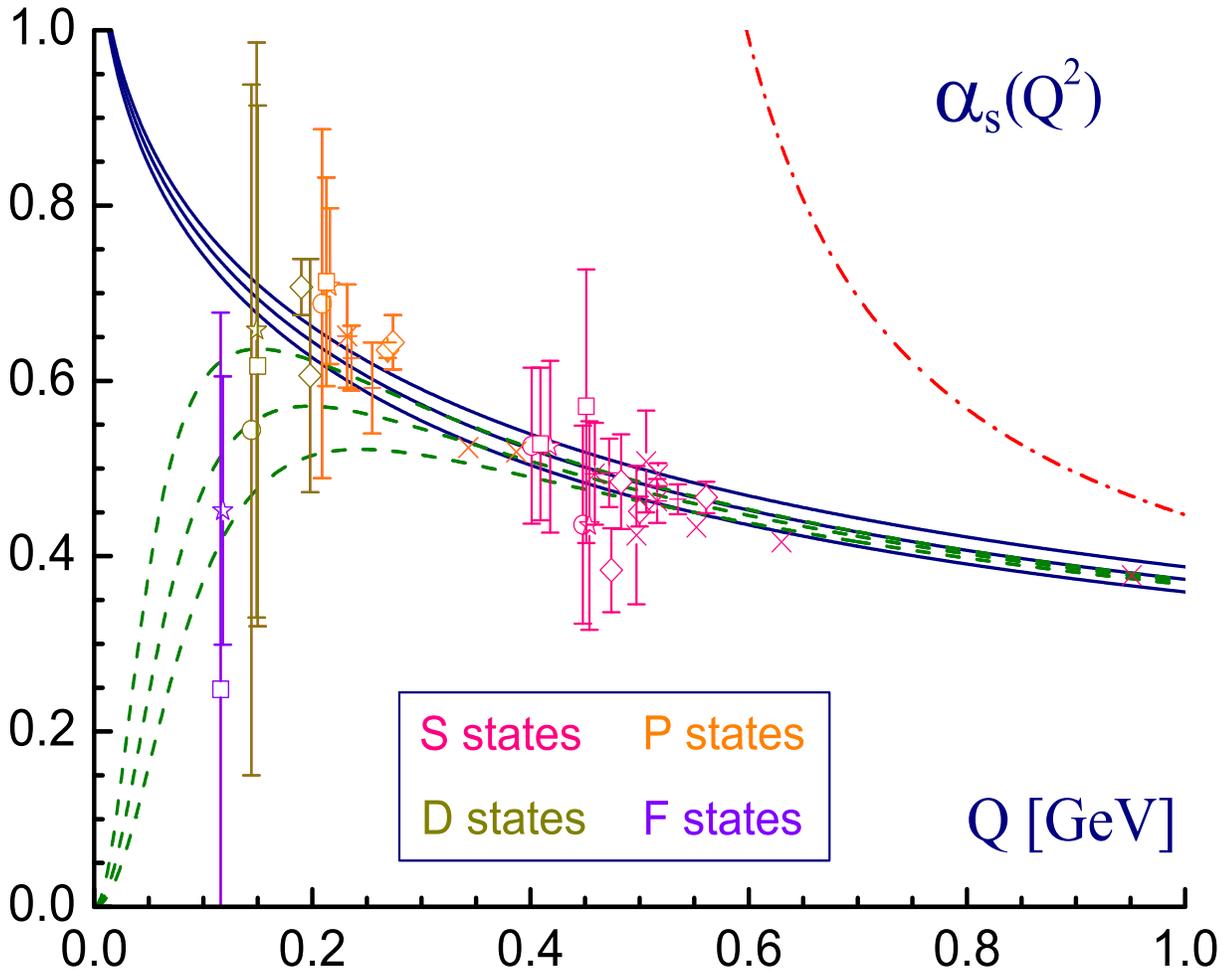}
\vspace{1.1truecm}
\caption{{\footnotesize Extracted values of $\alpha_{s}^{\rm exp}(Q^2)$
against the 3-loop APT coupling~(\ref{AAPT}) with
$\Lambda^{(3)}_{n_f=3}=(417\pm 42)\,$MeV (solid), and its
perturbative counterpart (dot-dashed). The ``massive'' 1-loop APT
coupling ($n=1$ in~(\ref{AMAPT})) refers to $\Lambda^{(1,{\rm
eff})}_{n_f=3}=204\,$MeV and $m\ind{}{eff}=(38 \pm 10)\,$MeV
(dashed). Circles, stars and squares refer respectively to $q\bar
q\,$, $s\bar s\,$ and $q\bar s\,$ with $q=u,d\,$, diamonds and
crosses to $c\bar c\,$ and $b\bar b\,$; asterisks stay for $q\bar
c\,$ and $q\bar b\,$, whereas plus signs for $s\bar c\,$ and $s\bar
b\,$.}}
\lb{low}
\end{figure}

\end{document}